# Overdamped van der Waals Josephson junctions by area engineering

Annu Anns Sunny, Harshit Choubey, Ankit Khola, Sreevidya Narayanan, Rajat Bharadwaj, Parvathy Gireesan & Madhu Thalakulam [*]

School of Physics, IISER Thiruvananthapuram, Kerala, India 695551

**Abstract**

Van der Waals (vW) Josephson junctions (JJs) realized by stacking materials such as few-layered $NbSe_2$, offers a new landscape to realize superconducting quantum devices with superior properties owing to its crystalline nature and defect-free junctions. For quantum technology, overdamped JJs are highly sought-after, whose realization demands precise control of junction capacitance by engineering the junction area using microfabrication techniques. $NbSe_2$ is highly reactive and susceptible to damage during microfabrication processes. In this manuscript, we demonstrate both underdamped and overdamped $NbSe_2$-$NbSe_2$ JJs by controlling the junction area. We devise a minimally invasive microfabrication procedure, post-junction formation, to precisely control the junction area. The McCumber parameter characterizing the damping is extracted from the electrical transport measurements down to 130 mK. The results show that our sample fabrication recipe has preserved the material qualities and paved the way for the realization of scalable JJ devices on $NbSe_2$ and similar systems.

---

[*] madhu@iisertvm.ac.in

**Introduction**

The Josephson junctions (JJ) are the building blocks of superconducting quantum circuits such as superconducting qubits[1], RSFQ digital electronics[2], Josephson voltage standards[3], and SQUIDs[4,5]. While aluminium (Al) or niobium (Nb) makes up the superconducting layers, the insulating layer mostly comprises $AlO_x$. Ease of fabrication and the presence of a robust native oxide make Al-based JJs a preferred choice over the Niobium ones[6]. Al–$AlO_x$–Al tunnel junctions produced using the double-angle evaporation process[7] have been widely explored for superconducting qubits[8] and quantum electrical metrology circuits[3]. Coherence time, the duration over which the quantum mechanical behavior of the system is preserved, is probably the single and most important parameter defining the quality of the system for quantum technologies, especially for quantum information processing[9]. The amorphous nature of the $AlO_x$ layer, structural defects such as dangling bonds at the metal/oxide interface[10], uneven oxide thickness[4], surface inhomogeneity, etc., are detrimental to the coherence times and device performance. One of the leading causes of energy dissipation, noise, and decoherence in superconducting electric circuits has been identified as the two-level systems owing to the reasons mentioned above in the $AlO_x$ layer[10–16]. Dielectric loss in $AlO_x$ is another reason for the reduced coherence time[17–19]. For effective Josephson coupling, the thickness of the oxide layer needs to be ~ 1 nm. A crystalline superconductor and an atomically thin and lossless dielectric material perfectly flat across the interface are the perfect candidates to overcome the above-mentioned issues. By definition, the surface of three-dimensional systems is reconstructed, and the physical deposition of Al or Nb results in amorphous or polycrystalline superconducting films that are not free of defects and disorders. Any third material, such as the dielectric medium, can provide interfacial defects, lattice mismatch, dangling bonds, dielectric loss, etc., making the system dissipative. Oxide films, by definition, are defective; growing defect-free atomically thin oxide layers is practically impossible.

In contrast, inherently two-dimensional (2D) systems, specifically the van der Waals (vW) superconducting materials, have the potential to circumvent all these issues together. They offer pristine, atomically flat, low impurity density surfaces with clean intefaces[20]. Among the vW systems, the transition-metal dichalcogenides (TMDC) boast an array of materials covering a spectrum of electrical properties, $NbSe_2$ [21,22], $NbS_2$[23], $TaSe_2$[24], and $TaS_2$[25,26] being the popular ones. Survival of superconductivity down to the mono-layer limit in these systems has been demonstrated[23,27]. Superconducting tunnelling devices consisting of $MoS_2$[28], $WSe_2$[29], hBN[30], and graphene[31] tunnel barriers have been explored in these systems. The prospect of using the vW gap as the tunnel junction altogether eliminates the need for a third dielectric medium to form the JJ and minimizes the dielectric loss. In this regard, vW JJs are fabricated by micromechanical exfoliation and vertical stacking of $NbSe_2$[32–34]. Reports on high-quality vW JJs on platforms such as $NbSe_2$, $NbS_2$, and $TaS_2$ claim improved transport properties and are promising candidates for quantum technologies.[32,35,36]

For technological applications such as programmable voltage standards[37] and RSFQ electronic circuits[38], one needs overdamped JJs. Josephson junctions are understood by the Resistively Capacitive Shunt Josephson model (RCSJ model)[39,40]. The damping characteristics of the system are characterized by the Stewart-McCumber parameter, $\beta_c = \frac{2eI_sR_n^2C}{\hbar}$ where $R_n$ is normal state resistance determined from the linear portion of the Current-Voltage Characteristics (I-V), C is the junction capacitance, $\hbar$ the Plank's constant and $e$ is the elementary charge[40]. $\beta_c < 1$ corresponds to overdamped systems, while $\beta_c$ above unity corresponds to underdamped cases [41,42]. The capacitive and resistive parts of the JJ control the damping levels of the system. Since the vW gap puts a lower limit on the junction thickness, one needs to control the junction area to engineer $\beta_c$ and the damping. However, the junction area created by deterministically stacking the flakes is not controllable, and one must resort to microfabrication techniques to control it. In contrast, TMDC superconductors are highly unstable, prone to oxidation, and defective even when exposed to ambient temperature and visible and UV radiation,[22,43] seriously limiting the prospects of conventional microfabrication process.

In this work, we focus on three micromechanically stacked NbSe$_2$-NbSe$_2$ vW JJ devices and demonstrate that controlling the junction area allows one to realize both underdamped and overdamped junctions to suit various technology needs. In addition, we also realize overdamped JJ in a scalable manner by controlling the junction area with a combination of electron-beam (e-beam) lithography followed by reactive ion etching (RIE). The entire fabrication process has been optimized to be minimally invasive and non-destructive. In-depth electrical transport measurements down to a temperature of 130 mK show that the device fabrication procedures adopted did not deteriorate the material qualities and device parameters and are comparable to those not subjected to any post-junction-formation treatments. From the transport measurements, we find that JJs with a larger area show an underdamped behavior while JJs with a lower area clearly show an overdamped characteristic. In-plane magnetotransport measurements reveal the Fraunhofer pattern in the critical current, a touchstone phenomenon of JJs.

The Josephson junction devices are realized by successive transfer of two few-layered NbSe$_2$ flakes onto a Si/SiO$_2$ or intrinsic Silicon substrate, maintaining a small overlap between the flakes. The vW gap between the flakes defines the junction. The flakes are mechanically exfoliated from commercially procured high-quality single crystals (HQ graphene) and accurately placed onto the lithographically defined gold source and drain electrodes using a residue-free PDMS-assisted transfer process in a home-made transfer setup having a lateral precision of ~ 2 μm[44]. Immediately after transferring the second flake, the entire device is encapsulated in a large area of the h-BN flake to limit subsequent exposure to the ambient atmosphere. We focus on three devices in this work, one underdamped and two overdamped JJs; for JJ1(larger junction area, underdamped) and JJ2 (lower junction area, overdamped) the junction area is defined by the overlap between the pristine flakes, and for the third one, JJ3( lower junction area, overdamped the junction area is precisely defined by e-beam lithography followed by RIE process of the overlapped region. We employ a combination of PMMA 495 and PMMA 950 (micro-chem) e-beam resists with an optimized baking time and temperature of 3 min each at 120 degrees. Samples baked at higher temperatures exhibit higher resistance owing to thermal degradation of the NbSe$_2$ [see Supporting Information SI-1]. We use a mixture of SF$_6$:O$_2$ (30:18 sccm) at an RF power of 60 Watts to perform the RIE. The fabrication recipe was optimized on single flakes before proceeding with etching the junction samples. To minimize any damage to the sample, we refrain from acquiring any SEM images until the measurements are completed. The measurements are carried out at high-vacuum using a 2 K closed cycle cryostat equipped with a 2 T solenoid magnet and a dilution refrigerator with a base temperature of 10 mK equipped with an 8 T superconducting magnet. All the transport data are acquired with a combination of low-noise source-measure units and lock-in amplifiers.

**Results and Discussion**

First, we focus our discussion on the NbSe$_2$-NbSe$_2$ JJ devices, which are not subject to any device fabrication processes or treatments post-junction formation. Fig.1 (a) shows the four-probe (4P) resistance versus temperature (R-T) down to a temperature of 4 K with a current bias of 10 µA applied across the junction for JJ1. The top left inset shows a scanning electron microscope (SEM) image of the device from which we calculate a junction area of 11.9 µm$^2$. The resistance of the device drops continuously down to a temperature of 7 K, revealing the metallic nature of the NbSe$_2$. On further reduction of temperature, the resistance shows a sharp drop to an immeasurably low value, signaling the transition to the superconducting state. Inset to Fig. 1(a) shows a magnified view of the R-T trace in the vicinity of the superconducting transition from which we extract a $T_c$ ~ 6.8 K for the sample, the temperature at which the resistance drops to 10% of the normal state resistance. Fig. 1 (b) shows the 4P R-T trace for JJ2 while a 10 µA applied across the junction. From the SEM image shown on the top left inset, we extract a junction area of 4.23 µm$^2$ for JJ2. Similar to JJ1, the sample shows a drop in the resistance down to a temperature of 9 K. Upon further reduction in the temperature, the sample undergoes a transition to the superconducting state. The lower-right inset shows a magnified view of the R-T trace in the vicinity of the superconducting state from which we extract a $T_c$ ~ 5. 01 K. We note here that the NbSe$_2$ flakes forming the junction in JJ2 are much thinner compared to that for JJ1, resulting in a lower $T_c$ [45].

The I−V characteristics for various temperatures for a few representative temperatures for JJ1 and JJ2 are shown in Fig. 1 (c) and (d), respectively. The insets in Fig. 1(c) and (d) show a magnified view of the I-V characteristics acquired in the forward and backward sweep directions in the 1st quadrant when the samples are superconducting at 4 K for JJ1 and at 4.38 K for JJ2. The I-V for JJ1, inset to Fig. 1 (c), shows a hysteretic behavior, a signature of underdamped JJs. The McCumber parameter $\beta_c = (\frac{4I_s}{\pi I_r})^2$, where $I_s$ and $I_r$ are the switching and re-trapping currents respectively[39]. The switching current is the critical current at which the junction transitions from a superconducting to a resistive state, while the re-trapping current is the lower critical current at which the junction returns to the superconducting state after being in the resistive state[39]. From the I-V characteristics [inset Fig. 1(c)], we extract $I_s$ ~ 0.77 mA, $I_r$ ~ 0.61 mA, and $\beta_c$ ~2.75, suggesting the junction in JJ1 is underdamped. From the $I_s$ and the junction area, we estimate a critical current density of 6750 Acm$^{-2}$ for JJ1. From the $\beta_c$ value, we obtain a junction capacitance of 0.14 pF corresponding to ~ 1.2 µFcm$^{-2}$; $R_n$ = 2.88 Ω is normal state resistance. Assuming the formation of a clean vW junction between the flakes and from the geometrical area of the junction estimated from the SEM image [Fig. 1 (a) inset], we extract a junction thickness $d$ ~ 0.74 nm, which matches very well with the vW gap between the flakes[32].

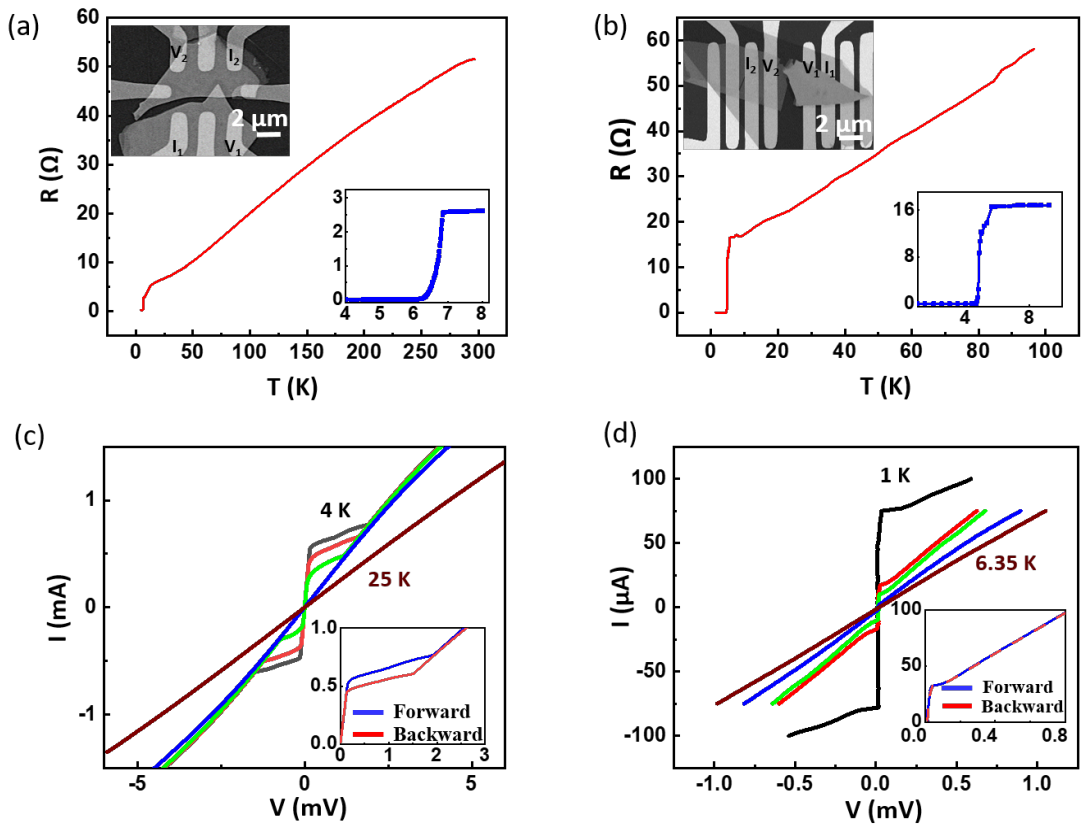

**Figure 1.** Temperature dependence of resistance (R-T) for (a) underdamped NbSe$_2$-NbSe$_2$ JJ (JJ1). Upper inset: SEM image of the sample. I$_1$ and I$_2$ are current and V$_1$ and V$_2$ are voltage probes. Lower inset: magnified view of the resistance for temperature range 4 K-8 K. (b) R-T trace for overdamped JJ (JJ2), between 300 K and 1 K. Upper inset: SEM image of the sample. I$_1$ and I$_2$ are current and V$_1$ and V$_2$ are voltage probes. Lower inset: magnified view of the resistance for temperature range 8 K-1 K. (c) I-V traces for the underdamped sample, JJ1, for temperatures between 4 K and 25 K. Inset: magnified view of the I-V trace at 4 K in the forward and backward sweep directions, showing the hysteretic behavior for JJ1. (d) I-V traces for the overdamped sample, JJ2, for temperatures between 1 K and 6.35 K. Inset: magnified view of the I-V trace at 4 K in the forward and backward sweep directions, showing the non-hysteretic nature of JJ2.

The JJ2 with a lower junction area exhibits non-hysteretic I-V characteristics in the superconducting state, as shown in the inset to Fig. 1 (d), a signature of overdamped JJs. From the I-V traces and the junction area, ~ 4.23 µm$^2$, obtained from the SEM image, we extract a critical current (current density) of 29 µA (685 Acm$^{-2}$), 64 µA (1512 Acm$^{-2}$), and 75 µA (1772 Acm$^{-2}$), at 4.25 K, 1.5 K, and 1 K respectively. Assuming a clean junction interface with a thickness of ~ 0.74 nm and from the junction area, we estimate a junction capacitance of ~ 50.3 fF, corresponding to ~ 1.2 µFcm$^{-2}$. Using the value of $I_c$= 75 µA (at 1 K) and $R_n$= 7.3 Ω, obtained from Fig. 1 (d), we calculate the McCumber parameter $\beta_c$ ~ 0.61, which is less than unity justifying the overdamped behavior of the JJ2.

The diagrams shown in Fig. 2 (a), from left to right, summarize the fabrication steps of the etched, overdamped sample, JJ3, in sequence. The rightmost panel shows a perspective SEM image of the final device, which is acquired post-measurements. Though the SEM image shows traces of the bottom flake, our measurement results, explained in the remaining part of the manuscript, confirm that the junction area is restricted by the etch. To minimize any damage to the sample, we restricted the etching time and power so that the top flake was completely etched to control the junction area. NbSe$_2$ is a highly unstable material; exposure to the plasma is sufficient to damage the flakes to restrict the supercurrent to the regions with no exposed plasma. We estimate a junction area of ~ 3.43 µm$^2$ from the SEM image.

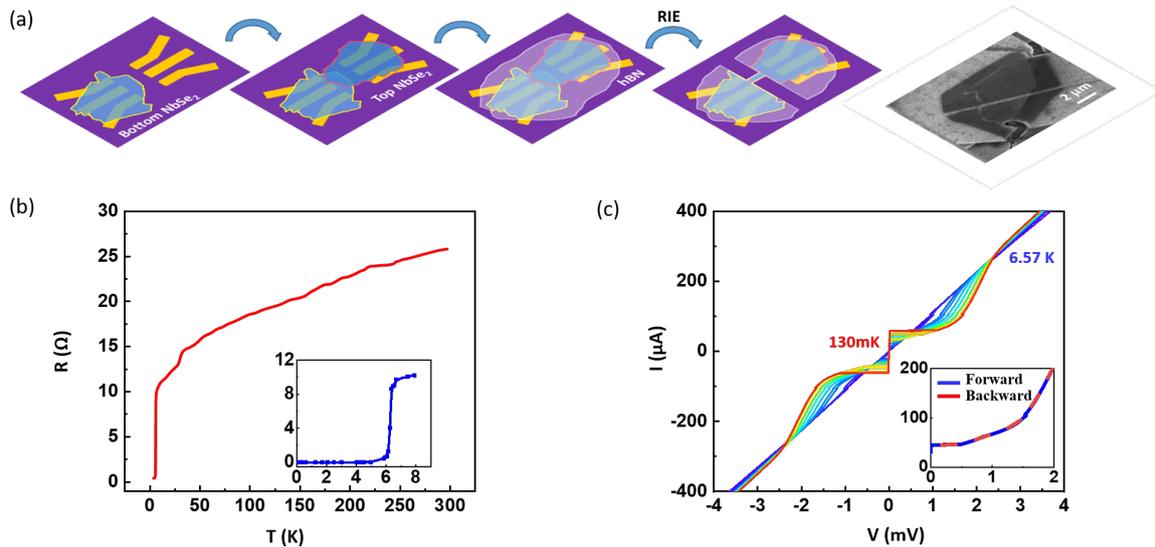

**Figure 2.** (a) Summary of steps adopted for etch-defined JJs (JJ3). The SEM image of the final device is shown in the rightmost panel. (b) R-T for temperatures between 300 K and 130 mK. Inset: magnified view of the R-T trace between 8 K and 130 mK. (c) I-V characteristics at different temperatures between 130 mK and 6.57 K. Inset: I-V characteristics at 4 K showing the non-hysteretic behavior of the overdamped JJ.

Fig. 2 (b) shows the R-T trace for JJ3. The resistance of the sample undergoes a continuous reduction as the temperature is lowered and drops to the noise level below a temperature of 7 K. The inset to Fig. 2 (b) shows a magnified view of the R-T trace below 8 K, from which we estimate a $T_c$ ~ 6.5 K. The $T_c$ value is similar to that of both the unetched JJ device shown in Fig.1 and the single pristine flake shown in Supporting Information, SI-2. In addition, the nature of the R-T plot we observe for the etch-defined sample is similar to that of the unetched JJ samples. All these point to the fact that our lithographic and etching processes have not induced any significant damage to the device or any detrimental effect on the basic characteristics of the junction. The flakes forming JJ3 are visibly thinner than those we used to form JJ1 and the pristine single flake SI-2, which we believe caused a slight drop in the $T_c$ from ~ 6.75 K to ~ 6.5 K.

Fig. 2 (c) shows I-V characteristics of JJ3 acquired at different temperatures starting from 130 mK through 10 K. The inset to Fig. 2 (c) shows I-V traces obtained at 130 mK for forward (solid) and backward (broken) sweep directions. The absence of hysteresis, similar to that of JJ2 with a similar junction area, confirms the overdamped nature of the JJ. From the junction area, ~ 3.43 µm$^2$, estimated from the SEM image, and thickness of ~ 0.74 nm (the junction was formed with a procedure similar to that of the unetched ones), we estimate a junction capacitance of ~ 40.83 fF, which corresponds to 1.2 µF cm$^{-2}$. The McCumber parameter $\beta_c \approx 0.608$, less than unity, obtained from the critical current $I_c$= 60 µA and $R_n$= 9.03 Ω reiterate the overdamped behavior of the JJ. Also note that another voltage jump at 259 µA can be attributed to the breaking of superconductivity in NbSe2 flakes, corresponding to the normal resistance of 9.03 Ω. A curvature in the I-V characteristics has been observed and can be due to quasiparticle currents, as discussed elsewhere[33].

A plot of the differential conductance versus voltage across the junction for various temperatures, extracted from the I-V traces in Fig 2 (c), is shown in Fig 3 (a). The peaks located ~ ± 2.1 meV mark the edge of the superconducting gap for the NbSe$_2$ flake, $2\Delta(T)$, at 130 mK. The broken lines are a guide to the eye, showing the evolution of the superconducting gap of the NbSe$_2$ flake versus temperature. NbSe$_2$ is regarded as a multiband superconductor consisting of a total of three bands[46]; two approximately cylindrical Fermi surfaces correspond to the Nb-derived bands centered on Γ and K points and a third Se-derived pocket around the Γ point. The intrinsic gap belongs to the bands associated with the Nb-derived one at the K-point, while smaller gaps arise due to the inter-band coupling between the other two bands[45]. From Fig. 3 (a), we identify the gap 2Δ~ ±2.1 meV as the intrinsic gap (at 130 mK), which exhibits nearly a BCS-like temperature dependence. The data also shows a smaller gap of $2\Delta_s \sim \pm 1\, meV$ (for 130 mK), which we believe is evolving from the quasiparticle inter-band coupling. The dotted line in Fig. 3 (a) is a guide to the eye representing the temperature dependence of this feature[33]. In addition, we also observe a resonant feature between the larger and smaller superconducting gaps, as denoted by the dash-dotted lines in Fig. 3 (a). Though the exact origin of these needs to be probed, we think the inter-band quasiparticle tunnelling across the junction can be a reason for these features [47].

Fig. 3 (b) shows the variation of the superconducting gap for JJ3 with temperature. The temperature-dependent energy gap is given by BCS theory: $\Delta(T) = \Delta(0) tanh\,[2.2\sqrt{(T_c/T) - 1}]$, where $\Delta(0)$ is the gap at absolute zero. The red trace in Fig. 3 (b) fits the preceding equation, from which we extract a superconducting gap $\Delta(0)$= 0.97 meV and $T_c$ = 6.72 K for JJ3. The extracted $\Delta(0)$ and $T_c$ values are in good agreement with the estimated ones from Fig. 1 (a), using BCS approximation; $\Delta(0) = 1.76 k_B T_c = 1.032$ meV. Inset to Fig. 3 (b) shows the critical current $I_c$ versus temperature for JJ3 obtained from the I-V traces shown in Fig. 2 (c). The temperature dependence of critical current is given by the Ambegaokar−Baratoff (AB) equation[48]

$$\frac{I_c(T)}{I_c(0)} = \frac{\Delta(T)}{\Delta(0)} \tanh\left(\frac{\Delta(T)}{2k_B T}\right)$$

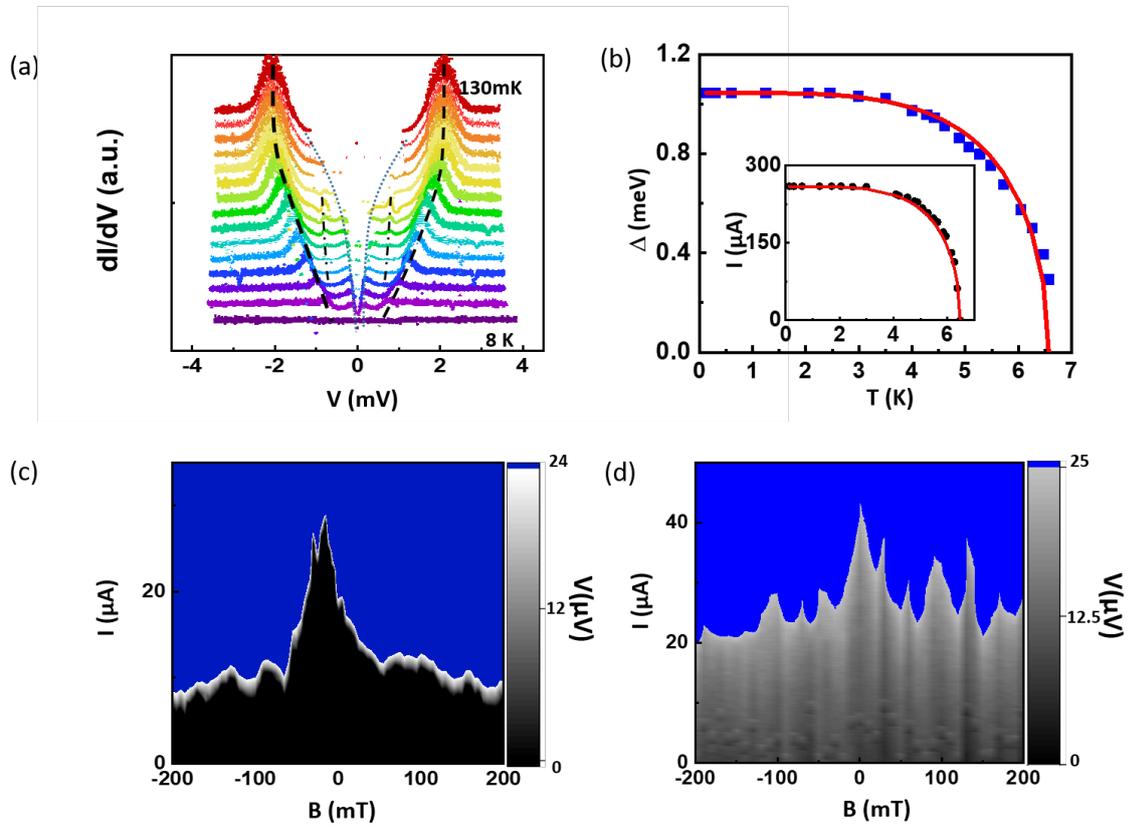

**Figure 3.** (a) Differential conductance plot for various temperatures for the etch-defined overdamped JJ(JJ3). Dashed line is used to represent the temperature dependence of superconducting gap $\Delta$ and dotted lines are used to represent temperature dependence of smaller superconducting gap $\Delta_s$. Dash-dotted lines highlights the inter-band quasiparticle tunnelling across the junction. (b)Temperature dependence of the superconducting gap $\Delta$ for JJ3. Red line fits to BCS theory. Inset: Temperature dependence of the critical current $I_c$ for JJ3. The red line fits to AB theory. (c) Contour plot of voltage across the junction as a function of bias current and external field measured at T = 4.3 K for JJ2. (d) Contour plot of voltage across the junction versus the bias current and external field measured at T = 4 K for JJ3. The Fraunhofer pattern in the voltage is a signature of JJ.

The red trace fits the AB equation and exhibits excellent agreement with the theory; we extract $I_c(0) = 259$ µA from the fit.

An in-plane field induces a phase shift in a direction normal to the field, and as a result, the junction critical current $I_c$ will display a Fraunhofer pattern as the magnetic field is varied [40]. Fig. 3 (c) and (d) show contour plots of voltage across the junction as a function of the bias current and the external field at 4 K and 4.35 K for samples JJ2 and JJ3, respectively. The periodic modulation in voltage across the junction against the magnetic field resembles the expected Fraunhofer pattern. We note here that the minima of $I_c(B)$ does not reach zero, possibly caused by flux focusing[49,50] and the uneven interface in the vicinity of the electrodes[36,51]. Any relative angle between the junction and the magnetic field caused while mounting the sample can also result in this behavior [52]. The presence of any perpendicular field components can lead to vortices in the system.

**Conclusions**

In conclusion, we realized vW Josephson junctions by micromechanical exfoliation and stacking of few-layered NbSe$_2$ flakes. We also demonstrated that the McCumber parameter defining the damping characteristics of the Josephson junctions can be tuned by controlling the junction area using a modified microfabrication routine such that the sample qualities are preserved. The samples subjected to fabrication processes exhibit junction characteristics comparable to those of pristine ones with similar areas. The procedures adopted in this manuscript can lead to a strategy for the realization of superconducting quantum circuits on highly crystalline two-dimensional systems in a scalable manner.

**Acknowledgments:** MT acknowledges SERB, Govt. of India, for the financial support under the grand CRG/2018/004213. AAS acknowledges UGC for Fellowship.

**Author contributions:** MT conceived the problem. AAS, HC, and AK fabricated the devices, AAS, HC, SN, PG, and RB performed the measurements, AAS & HC performed the data analysis, and AAS and MT co-wrote the manuscript.

**Supporting information: Overdamped van der Waals Josephson junctions by area engineering**

Annu Anns Sunny, Harshit Choubey, Ankit Khola, Sreevidya Narayanan, Rajat Bharadwaj, Parvathy Gireesan & Madhu Thalakulam[*]

School of Physics, IISER Thiruvananthapuram, Kerala, India 695551

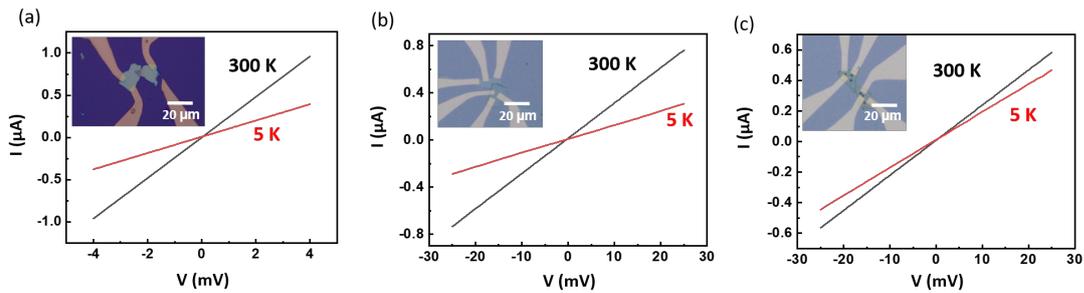

SI -1. I-V characteristics of annealed samples. (a) I-V characteristics at different temperature. Inset:optical image of sample 1. (b) I-V characteristics at different temperature. Inset:optical image of sample 2. (c) I-V characteristics at different temperature. Inset: optical image of sample 3.

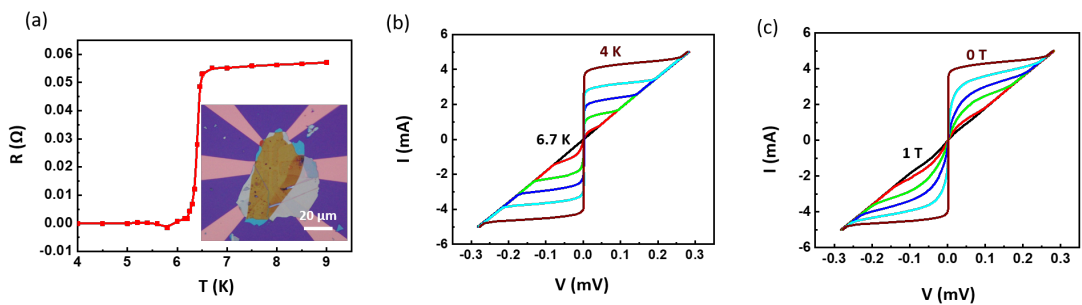

SI-2. Single flake transport measurement . (a) Temperature dependence of Resistance (9 K-4 K) Inset:optical image of the sample (b) I-V characteristics at different temperature (c) I-V characteristics at different perpendicular magnetic field at 4 K.

[*] madhu@iisertvm.ac.in